\documentclass[aps,pre,twocolumn,groupedaddress]{revtex4-1}
\usepackage{graphicx}
\usepackage{dcolumn}
\usepackage{bm}
\usepackage{amsmath}
\usepackage{amssymb}
\usepackage{color}

\begin{document}

\title{Membrane domain formation induced by binding/unbinding of curvature-inducing molecules onto both membrane surfaces}

\author{Hiroshi Noguchi}
\email[]{noguchi@issp.u-tokyo.ac.jp}
\affiliation{Institute for Solid State Physics, University of Tokyo, Kashiwa, Chiba 277-8581, Japan}


\begin{abstract}
The domain formation of curvature-inducing molecules, such as peripheral or transmembrane proteins and conical surfactants,
is studied in thermal equilibrium and nonequilibrium steady states using meshless membrane simulations.
These molecules can bind onto both surfaces of a bilayer membrane
and also move to the opposite leaflet by a flip--flop.
In symmetric conditions for the two leaflets, the membrane domains form checkerboard patterns 
in addition to stripe and spot patterns.
The unbound membrane stabilizes the vertices of the checkerboard.
In asymmetric conditions, the domains form kagome-lattice and thread-like domains.
In the nonequilibrium steady states, a flow of the binding molecules between the upper and lower solutions can occur via the flip--flop.
\end{abstract}

\maketitle

\section{Introduction}

In living cells, biomembranes consist of many types of lipids and proteins,
and their ratios vary depending on the types of cells and organelles.\cite{meer08,anto15,ingo14}
The biomembranes are laterally inhomogeneous and
microdomains called lipid rafts are formed to serve as a platform for protein functions and signal transactions.\cite{simo00,ling10}
Two leaflets of biomembranes are maintained with different lipid compositions via vesicle transport and flippases, etc.\cite{dale03,cont09}
Flippases and floppases move specific lipids in one direction (flip and flop, respectively) using ATP hydrolysis; while
scramblases move lipids in both directions, equilibrating the lipid distribution of two leaflets.
The biomembrane shapes are regulated by various types of curvature-inducing proteins;
 Bin/Amphiphysin/Rvs (BAR) superfamily proteins contain a crescent binding domain to bend the membrane along the domain axis,\cite{mcma05,suet14,joha15,itoh06,masu10,baum11,mim12a,fros08,adam15,snea19} while clathrin and coat protein complexes isotropically bend membranes and form spherical buds.\cite{joha15,bran13,hurl10,mcma11,kaks18}

Experimentally, lateral phase separation has been intensively studied for three-component lipid membranes,
in which liquid-ordered and liquid-disordered phases are formed.\cite{baum03,veat03,yana08,chri09}
Moreover, protein assemblies to generate spherical buds and tubules have been observed.
In theory and simulations, phase separation is induced by the line tension between different phases, bending-rigidity difference, and (isotropic or anisotropic) spontaneous-curvature difference.\cite{kuma99,akim07,gutl09,hu11,tani11,amaz13,acke15,hime15,rama18,nogu12a,nogu22a,nogu14,nogu15b,nogu16,nogu16a,nogu17,nogu17a,nogu19a,nogu22b,gout21} 
The separated phases exhibit circular and stripe patterns and induce membrane deformation such as spherical buds, straight bumps, and tubules.
However, in most cases, two phases have been considered. More complex domain patterns are expected in the presence of three or more phases.  Sunil Kumar et al. have theoretically studied a three-phase system in the weak-segregation limit and reported that square domains exist as metastable states\cite{kuma99}. Here, we will show that square domains with checkerboard alignment are formed in thermal equilibrium under a strong-segregation condition.

\begin{figure}[t]
\includegraphics[]{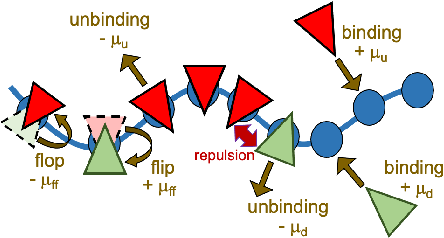}
\caption{
Schematic of the binding and unbinding of curvature-inducing proteins to a membrane.
Proteins (red and green triangles) have a finite spontaneous curvature. 
The proteins bind to the membrane  from upper and lower solutions with
binding chemical potentials $\mu_{\mathrm{u}}$ and $\mu_{\mathrm{d}}$, respectively.
The proteins move between the upper and lower membrane surfaces
with the flip--flop chemical potential $\mu_{\mathrm{ff}}$.
}
\label{fig:cart}
\end{figure}

In this study, we consider three types of membrane states (two types of the bound states and one unbound state) 
based on the membrane model used in our previous study,\cite{gout21}
in which proteins bind only onto one surface of the membrane.
The bound states have positive or negative spontaneous curvature with the same amplitude in the Canham--Helfrich energy.\cite{safr94,canh70,helf73}
That is, the same type of molecules bind onto the upper and lower membrane surfaces (see Fig.~\ref{fig:cart})
and the bound molecules are considered to be laterally isotropic.
Since two solutions above and below the membrane are separated,
the binding onto upper and lower surfaces can have different chemical potentials.

The transmembrane motion of the molecules between the upper and lower surfaces (flip--flop) is also considered (see Fig.~\ref{fig:cart}).
Although the flip--flop of phospholipids is extremely slow (hours or days),\cite{korn71}
amphiphilic molecules having small hydrophilic head groups such as 
cholesterols and diacylglycerols exhibit much faster flip--flop (the half lifetimes are less than a minute).\cite{cont09,hami03,stec02}
In living cells, enzymes (flippases, floppases, and scramblases) promote lipid transport between the two leaflets.
Proteins and peptides do not directly perform flip--flop but can move to the opposite surface via transmembrane pore formation\cite{guha19,sato06,fisc07} and with the help of translocon.\cite{hegd11,cyme15} 
In this study, we do not consider such intermediate states but consider symmetric bound states with respect to the membrane middle plane for simplicity.
In nonequilibrium steady states, the binding molecules can flow from the upper and lower solutions via flip--flop.
Previously, such a flow has been reported for vesicles, in which inner and outer solutions are different,
in experiments\cite{miel20,holl21} and simulations.\cite{naka18} 

In this study, we focus on general phase behaviors of binding molecules
so that we need not specify the type of binding molecules.
They can be curvature-inducing proteins, conical surfactants, or other amphiphilic molecules.
We examine several different conditions;
some of them are more easily realized by proteins in experiments, but the others by surfactants.
However, we call them upper proteins and lower proteins, respectively, for simplicity. 

The simulation model and method are described in Sec.~\ref{sec:method}.
The results are presented and discussed in Sec.~\ref{sec:results}.
The phase separation in thermal equilibrium  and nonequilibrium are described in Secs.~\ref{sec:eq} and \ref{sec:neq}, respectively.
Finally, a summary is presented in Sec.~\ref{sec:sum}.

\section{Simulation Model and Method}\label{sec:method}

In a meshless membrane model,
membrane particles self-assemble into  a one-layer sheet in a fluid phase.
Here, each membrane particle is a binding site and has three states: unbound (bare) membrane (denoted by n) and membrane bound by a protein
on the upper and lower surfaces (denoted by u and d), respectively (see Fig.~\ref{fig:cart}).
The proteins
are considered to have a high bending rigidity and positive spontaneous curvature
and exist with constant densities in both solutions above and below the membrane.
They bind to the upper and lower membrane surfaces with the binding chemical potential $\mu_{\mathrm{u}}$ and  $\mu_{\mathrm{d}}$, respectively.
The chemical potential for protein flip from upper to lower surfaces is  $\mu_{\mathrm{ff}}$ 
($-\mu_{\mathrm{ff}}$ for flop from the lower to upper surfaces).
In thermal equilibrium, $\mu_{\mathrm{d}}=\mu_{\mathrm{u}}+\mu_{\mathrm{ff}}$.
The different binding chemical potentials ($\mu_{\mathrm{u}} \neq \mu_{\mathrm{d}}$)
can be generated by the difference in bulk protein densities in the upper and lower solutions.
The flip--flop of the proteins are considered to occur spontaneously or actively via an enzyme.

The position and orientational vectors of the $i$-th particle are ${\bm{r}}_{i}$ and ${\bm{u}}_i$, respectively.
The membrane particles interact with each other via a potential $U=U_{\mathrm {rep}}+U_{\mathrm {att}}+U_{\mathrm {bend}}+U_{\mathrm {tilt}}+U_{\mathrm {pp}}$.
The potential $U_{\mathrm {rep}}$ is an excluded volume interaction with diameter $\sigma$ for all pairs of particles.
The solvent is implicitly accounted for by an effective attractive potential  as follows:
\begin{equation} \label{eq:U_att}
\frac{U_{\mathrm {att}}}{k_{\mathrm{B}}T} =  \frac{\varepsilon_{\mathrm{att}}}{4}\sum_{i} \ln[1+\exp\{-4(\rho_i-\rho^*)\}]- C,
\end{equation}
where  $\rho_i= \sum_{j \ne i} f_{\mathrm {cut}}(r_{i,j})$, $C$ is a constant,
 $\rho^*$ is the characteristic density with $\rho^*=7$,
and $k_{\mathrm{B}}T$ is the thermal energy.
 In this study, we employ $\varepsilon_{\mathrm{att}}=8$, as in our previous study,\cite{gout21} to prevent membrane rupture in a wider parameter range.
$f_{\mathrm {cut}}(r)$ is a $C^{\infty}$ cutoff function\cite{nogu06}
 and $r_{i,j}=|{\bf r}_{i,j}|$ with ${\bf r}_{i,j}={\bf r}_{i}-{\bf r}_j$:
\begin{equation} \label{eq:cutoff}
f_{\mathrm {cut}}(r)=
\exp\Big\{b\Big(1+\frac{1}{(r/r_{\mathrm {cut}})^n -1}\Big)\Big\}\Theta(r_{\mathrm {cut}}-r),
\end{equation}
where $\Theta(x)$ is the unit step function, 
 $n=6$, $b=\ln(2) \{(r_{\mathrm {cut}}/r_{\mathrm {att}})^n-1\}$,
$r_{\mathrm {att}}= 1.9\sigma$, and $r_{\mathrm {cut}}=2.4\sigma$.
The bending and tilt potentials
are given by 
\begin{eqnarray} \label{eq:ubend}
\frac{U_{\mathrm {bend}}}{k_{\mathrm{B}}T} &=& \frac{k_{\mathrm {bend}}}{2} \sum_{i<j} ({\bm{u}}_{i} - {\bm{u}}_{j} - C_{\mathrm {bd}} \hat{\bm{r}}_{i,j} )^2 w_{\mathrm {cv}}(r_{i,j}), \\
\frac{U_{\mathrm {tilt}}}{k_{\mathrm{B}}T} &=& \frac{k_{\mathrm{tilt}}}{2} \sum_{i<j} [ ( {\bm{u}}_{i}\cdot \hat{\bm{r}}_{i,j})^2
 + ({\bm{u}}_{j}\cdot \hat{\bm{r}}_{i,j})^2  ] w_{\mathrm {cv}}(r_{i,j}),\  
\end{eqnarray}
where 
 $\hat{\bm{r}}_{i,j}={\bm{r}}_{i,j}/r_{i,j}$ and $w_{\mathrm {cv}}(r_{i,j})$ is a weight function. 
The spontaneous curvature is given by $C_0 = C_{\mathrm {bd}}/2\sigma$.\cite{shib11} 
The details of a bare (unbound) membrane are described in Ref.~\citenum{shib11}.

For the unbound and bound membrane particles, the parameter values in Ref.~\citenum{gout21} are used in this study.
For the unbound particles,
$C_0=0$ and $k_{\mathrm {bend}}=k_{\mathrm{tilt}}=10$,
and for the bound particles, $C_0\sigma= \pm 0.1$ and
$k_{\mathrm {bend}}=k_{\mathrm{tilt}}=80$.
For a pair of neighboring bound and unbound particles,
the mean value ($k_{\mathrm {bend}}=k_{\mathrm{tilt}}=45$) are used.
The mean values are also used for the spontaneous curvature ($C_0\sigma= 0.05$, $-0.05$, and $0$ for u--n, d--n, and u--d pairs, respectively).
The bending rigidities of the unbound and bound membrane are $\kappa_{\mathrm {n}}/k_{\mathrm{B}}T=16 \pm 1$ and $\kappa_{\mathrm {u}}/k_{\mathrm{B}}T=\kappa_{\mathrm {d}}/k_{\mathrm{B}}T=144 \pm 7$, respectively
(calculated from the thermal undulation of a flat membrane~\cite{shib11}).
The saddle-splay modulus $\bar{\kappa}$ is proportional to $\kappa$, as $\bar{\kappa}/\kappa=-0.9\pm 0.1$.\cite{nogu19}
Note that the $\bar{\kappa}$ difference also affect the protein binding.\cite{nogu22a,nogu21a}
The different types of bound sites (u and d) have a repulsive interaction, by 
\begin{equation} \label{eq:cutoff}
\frac{U_{\mathrm {pp}}}{k_{\mathrm{B}}T} = \varepsilon_{\mathrm {pp}} \sum_{i\in u, j\in d} \exp\big( \frac{1}{r_{ij}/r_{\mathrm {cut}} -1} + b_0 \big)\Theta(r_{\mathrm {cut}}-r),
\end{equation}
where $b_0= r_{\mathrm {cut}}/(r_{\mathrm {cut}}-\sigma)$.\cite{nogu12a}
At $r_{ij}=\sigma$, the potential height is $\varepsilon_{\mathrm {pp}}k_{\mathrm{B}}T$.
Unless  otherwise specified, $\varepsilon_{\mathrm {pp}}=2$ is used.
Although the phase separation can be induced by the attraction between the same types of proteins,
we choose this repulsion to reproduce the phase behavior in our previous study\cite{gout21} in the case that the proteins bind only onto one surface of the membrane.
Thus, the unbound and bound sites are more easily mixed than the upper and lower proteins.

\begin{figure}[t]
\includegraphics[]{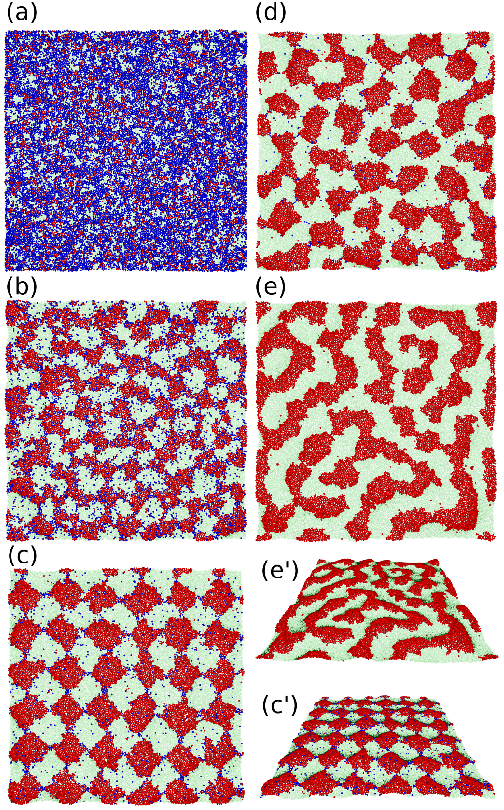}
\caption{
Snapshots of the membranes at up--down symmetrical conditions ($\mu_{\mathrm{u}}=\mu_{\mathrm{d}}$ and $\mu_{\mathrm{ff}}=0$) 
for $\varepsilon_{\mathrm {pp}}=2$ in thermal equilibrium.
(a) $\mu_{\mathrm{u}}/k_{\mathrm{B}}T=4$. (b) $\mu_{\mathrm{u}}/k_{\mathrm{B}}T=6$. (c) $\mu_{\mathrm{u}}/k_{\mathrm{B}}T=7.5$. 
(d) $\mu_{\mathrm{u}}/k_{\mathrm{B}}T=9$. (e) $\mu_{\mathrm{u}}/k_{\mathrm{B}}T=20$.
Red and light green spheres represent the upper and lower proteins, respectively.
Blue spheres represent unbound membrane particles.
Bird's-eye views of the membranes of (c) and (e) are also shown in (c') and (e'), respectively.
}
\label{fig:snapeqsym}
\end{figure}

\begin{figure}[t]
\includegraphics[]{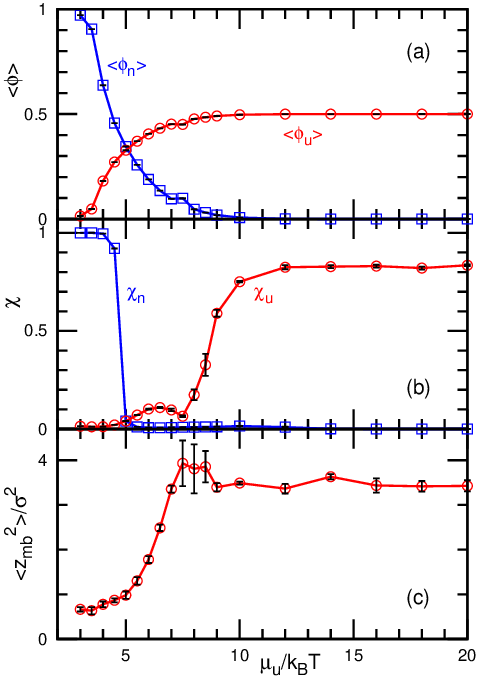}
\caption{
Dependence on the binding chemical potential at up--down symmetrical conditions ($\mu_{\mathrm{u}}=\mu_{\mathrm{d}}$ and $\mu_{\mathrm{ff}}=0$)
for $\varepsilon_{\mathrm {pp}}=2$ in thermal equilibrium.
(a) Density of bound proteins $\phi_{\mathrm{u}}=\phi_{\mathrm{d}}$ and unbound membrane $\phi_{\mathrm{n}}$.
(b) Mean cluster sizes normalized by the total number of each type of particle:
for the upper proteins, $\chi_{\mathrm{u}}=\langle N_{\mathrm{ u,cl}}\rangle/\langle N_{\mathrm{u}}\rangle$ and  
for the unbound membrane particles $\chi_{\mathrm {n}}=\langle N_{\mathrm {n,cl}}\rangle/\langle N_{\mathrm{n}}\rangle$. 
(c) Membrane vertical span $z_{\mathrm {mb}}$.
}
\label{fig:symmu}
\end{figure}

The membrane consisting of $25~600$ membrane particles is simulated under periodic boundary conditions.
The $N\gamma T$ ensemble is used, where $N$ is the total number of particles and $\gamma$ is the surface tension,
so that the projected area of the square membrane thermally fluctuates.\cite{fell95,nogu12}
Here, we use a tension of $\gamma=0.5k_{\mathrm{B}}T/\sigma^2$, which is sufficiently large to prevent the budding of a bound domain.\cite{gout21}
This tension corresponds to an average tension $\approx  0.02$\,mN/m, well below usual lysis tension ($1$--$25$\,mN/m),\cite{evan00,evan03,ly04}
($\sigma\approx 10$\,nm and $k_{\mathrm{B}}T\approx 4\times10^{-21}$\,J).
The structures of tensionless membranes ($\gamma=0$) are briefly described in Appendix~\ref{app1}.
The motion of the particle position ${\bf r}_{i}$ and 
orientation ${\bf u}_{i}$ are given by underdamped Langevin equations,
which are integrated by the leapfrog algorithm\cite{alle87,nogu11}
with $\Delta t=0.002\tau$. The time unit is $\tau= \sigma^2/D_0$,
where $D_0$ is the diffusion coefficient of the free membrane particles.

Two types of states [binding/unbinding (u vs. n and d vs. n) and flip--flop (u vs. d)] 
are stochastically switched by a Metropolis Monte Carlo (MC) procedure with the acceptance rate $p_{\mathrm {acpt}}$:
\begin{equation}\label{eq:Metro}
p_{\mathrm {acpt}} = \left\{
\begin{array}{ll}
\exp(\pm \Delta H/k_{\mathrm{B}}T ) &{\mathrm {if\ }} \pm \Delta H <0, \\
  1 &{\mathrm {otherwise}},
\end{array}
\right.
\end{equation}
where the $+$ and $-$ signs refer to the forward and backward processes, respectively.
Here, $\Delta H= \Delta U - \mu_{\alpha}$ where $\Delta U$ is the energy difference between the two states
and $\mu_{\alpha}$ is the chemical potential of each process.
Thermal equilibrium states are not dependent on the frequency of these MC processes.
Conversely, in nonequilibrium, steady states can be varied by this frequency.
In this study, we perform these MC processes every $\tau_{\mathrm {MC}}=0.01\tau$
with one MC trial per particle for flip--flop and with $0.01$ trial per particle for binding/unbinding.
The dependences on the rates of the flip--flop and binding/unbinding MC trials are described in Appendix~\ref{app2}.
 
To characterize various phases, mean cluster sizes are calculated.
Two sites are considered to belong to the same cluster
when the distance between them
is less than $r_{\mathrm {att}}$. 
The mean size of the clusters is given by
$N_{\mathrm {cl}}= (\sum_{i_{\mathrm {cl}}=1}^{N_{\alpha}} i_{\mathrm{cl}}^2 n^{{\mathrm {cl}}}_i)/N_{\alpha}$,
where $n^{{\mathrm {cl}}}_i$  is the number of clusters of size $i_{\mathrm{cl}}$ 
and $N_{\alpha}$ is the total number of each state (bound (u and d) and unbound states (n)).
The mean cluster size of  each state
is normalized by the total number, e.g., $\chi_{\mathrm{u}}=\langle N_{\mathrm{ u,cl}}\rangle/\langle N_{\mathrm{u}}\rangle$. 
A large percolated cluster results in $\chi \simeq 1$.
The vertical span of the membrane is calculated from 
the membrane height variance as 
$z_{\mathrm {mb}}^2=\sum_{i}^{N} (z_i-z_{\mathrm G})^2/N$,
where $z_{\mathrm G}=\sum_{i}^{N} z_i/N$. 
Error bars are calculated from three or more independent runs.

\section{Results and Discussion}\label{sec:results}

\subsection{Phase Separation in Thermal Equilibrium}\label{sec:eq}

\begin{figure}[t]
\includegraphics[]{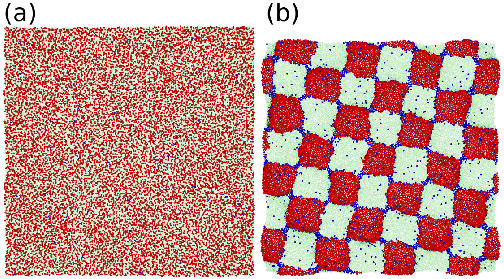}
\includegraphics[]{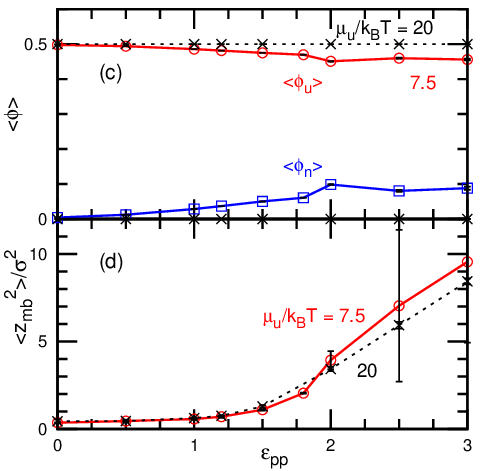}
\caption{
Dependence on the repulsion strength $\varepsilon_{\mathrm {pp}}$ at up--down symmetrical conditions ($\mu_{\mathrm{u}}=\mu_{\mathrm{d}}$ and $\mu_{\mathrm{ff}}=0$) in thermal equilibrium.
(a), (b) Snapshots for (a) $\varepsilon_{\mathrm {pp}}=0$ and (b) $\varepsilon_{\mathrm {pp}}=3$ at $\mu_{\mathrm{u}}/k_{\mathrm{B}}T=7.5$.
(c) Density of bound proteins $\phi_{\mathrm{u}}=\phi_{\mathrm{d}}$ and unbound membrane $\phi_{\mathrm{n}}$.
(d) Membrane vertical span $z_{\mathrm {mb}}$.
The solid and dashed lines represent the data for $\mu_{\mathrm{u}}/k_{\mathrm{B}}T=7.5$ and $20$, respectively.
}
\label{fig:symr}
\end{figure}

First, we describe the protein distribution under up--down symmetrical conditions in thermal equilibrium,
i.e., $\mu_{\mathrm{u}}=\mu_{\mathrm{d}}$ and $\mu_{\mathrm{ff}}=0$.
The proteins bind the upper and lower surfaces with the same amount, as shown in Figs.~\ref{fig:snapeqsym}--\ref{fig:symr}.
As the binding chemical potentials ($\mu_{\mathrm{u}}=\mu_{\mathrm{d}}$) increase,
the protein densities ($\phi_{\mathrm{u}}=\phi_{\mathrm{d}}$) increase,
leading to the formation of protein domains (see Figs.~\ref{fig:snapeqsym} and \ref{fig:symmu}(a)).
Here, the densities are normalized as $\phi_{\mathrm{u}}+\phi_{\mathrm{d}}+\phi_{\mathrm{n}}=1$.
At $\mu_{\mathrm{u}}/k_{\mathrm{B}}T \gtrsim 10$,
the membrane is almost completely covered by the proteins,
and the proteins form stripe-shaped domains (see Fig.~\ref{fig:snapeqsym}(e)).
Most bound membranes belong to the largest domain (see Fig.~\ref{fig:symmu}(b)),
and the domains are percolated.
Since the membranes bound by proteins on the upper and lower surfaces 
bend in the positive and negative directions, respectively,
the membrane exhibits a bumped shape  (see Figs.~\ref{fig:snapeqsym}(e') and \ref{fig:symmu}(c)).

Interestingly, at an intermediate condition,
the membrane forms a checkerboard pattern, as shown in Fig.~\ref{fig:snapeqsym}(c) and (c').
The unbound membranes exist at the vertices of the checkerboard,
in which two domains of upper (lower) proteins are contacted.
The angle of the domain boundaries is $90^{\circ}$ owing to the symmetry.
The bound domains prefer to form a spherical-cap shape, alternatively in the upward and downward directions,
but the region of the vertex of the checkerboard deviates from it.
The unbound domain fills this region and  stabilizes it.
Although the stripe-shaped domains are often observed in membranes,
checkerboard patterns have not been previously reported in thermal equilibrium to the best of our knowledge.

\begin{figure}[t]
\includegraphics[width=7.8cm]{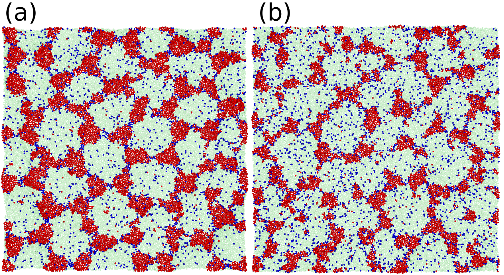}
\includegraphics[width=7.6cm]{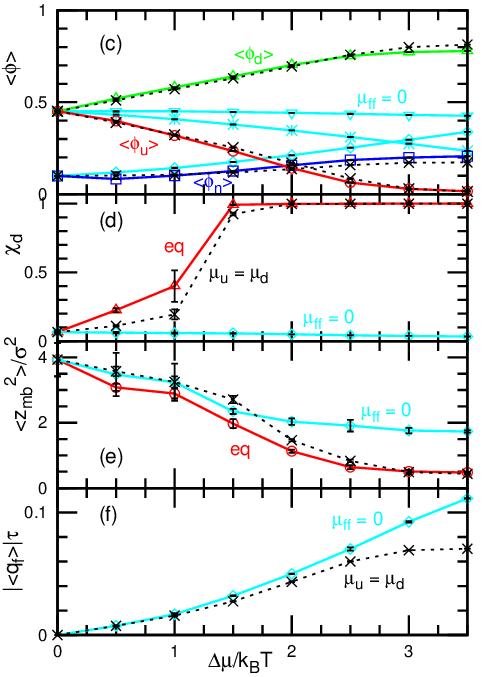}
\caption{
Protein binding at asymmetric conditions
with $\mu_{\mathrm{d}}/k_{\mathrm{B}}T=7.5$ and $\varepsilon_{\mathrm {pp}}=2$.
 $\mu_{\mathrm{u}}$ and/or $\mu_{\mathrm{ff}}$ vary.
In thermal equilibrium (labeled eq), $\Delta\mu =\mu_{\mathrm{ff}}= \mu_{\mathrm{d}}-\mu_{\mathrm{u}}$.
In nonequilibrium, two conditions, $\Delta\mu = \mu_{\mathrm{ff}}$ with $\mu_{\mathrm{u}}=\mu_{\mathrm{d}}$ and $\Delta\mu = \mu_{\mathrm{d}}-\mu_{\mathrm{u}}$ with $\mu_{\mathrm{ff}}=0$, are plotted.
(a), (b) Snapshots for (a) $\mu_{\mathrm{ff}}/k_{\mathrm{B}}T=1$ and (b)  $\mu_{\mathrm{ff}}/k_{\mathrm{B}}T=1.5$ in the thermal equilibrium condition.
(c) Densities of upper proteins $\phi_{\mathrm{u}}$, lower proteins $\phi_{\mathrm {d}}$,
and unbound membrane $\phi_{\mathrm {n}}$.
(d) Normalized mean cluster sizes $\chi_{\mathrm{u}}$ and $\chi_{\mathrm{d}}$.
(e) Membrane vertical span $z_{\mathrm {mb}}$.
(f) Flow rate $q_{\mathrm {f}}$ of the proteins between the upper and lower solutions.
The red and cyan solid lines represent the data in thermal equilibrium and in nonequilibrium with $\mu_{\mathrm{ff}}=0$, respectively.
The black dashed lines represent the data at $\mu_{\mathrm{u}}=\mu_{\mathrm{d}}$.
}
\label{fig:cp75asym}
\end{figure}

\begin{figure}[t]
\includegraphics[]{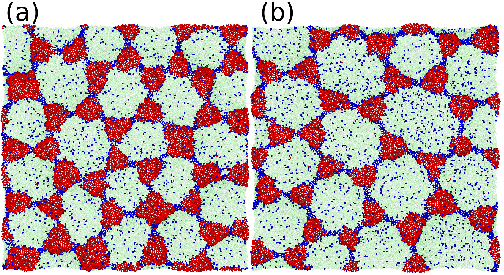}
\caption{
Snapshots at asymmetric conditions
with $\mu_{\mathrm{d}}/k_{\mathrm{B}}T=7.5$ and $\varepsilon_{\mathrm {pp}}=3$
in thermal equilibrium ($\mu_{\mathrm{ff}}= \mu_{\mathrm{d}}-\mu_{\mathrm{u}}$).
(a) $\mu_{\mathrm{ff}}/k_{\mathrm{B}}T=1$ and (b)  $\mu_{\mathrm{ff}}/k_{\mathrm{B}}T=1.5$.
}
\label{fig:cp75ar3}
\end{figure}

\begin{figure}[t]
\includegraphics[]{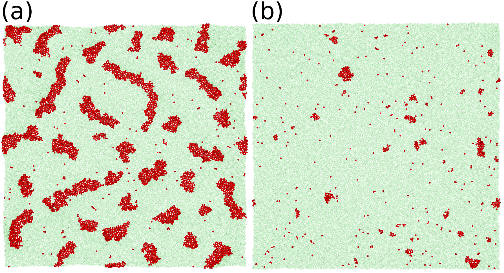}
\includegraphics[]{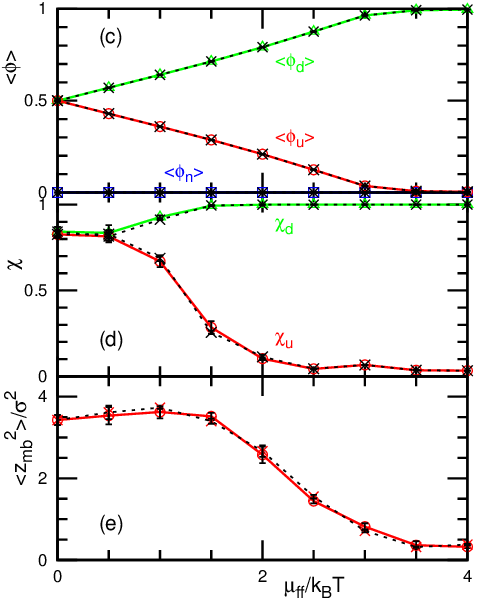}
\caption{
Dependence on the flip-flop chemical potential $\mu_{\mathrm{ff}}$
at $\mu_{\mathrm{d}}/k_{\mathrm{B}}T=20$ and $\varepsilon_{\mathrm {pp}}=2$.
(a), (b) Snapshots for (a) $\mu_{\mathrm{ff}}/k_{\mathrm{B}}T=2$ and (b)  $\mu_{\mathrm{ff}}/k_{\mathrm{B}}T=3$ in thermal equilibrium ($\mu_{\mathrm{u}}=\mu_{\mathrm{d}}-\mu_{\mathrm{ff}}$).
(c) Densities of upper proteins $\phi_{\mathrm{u}}$, lower proteins $\phi_{\mathrm {d}}$,
and unbound membrane $\phi_{\mathrm {n}}$.
(d) Normalized mean cluster sizes $\chi_{\mathrm{u}}$ and $\chi_{\mathrm{d}}$.
(e) Membrane vertical span $z_{\mathrm {mb}}$.
The solid and dashed lines represent the data for thermal equilibrium ($\mu_{\mathrm{u}}=\mu_{\mathrm{d}}-\mu_{\mathrm{ff}}$)
and nonequilibrium ($\mu_{\mathrm{u}}=\mu_{\mathrm{d}}$), respectively.
}
\label{fig:cp20asym}
\end{figure}

The phase separation between the upper and lower proteins
is controlled by the repulsion strength $\varepsilon_{\mathrm {pp}}$ (see Fig.~\ref{fig:symr}).
At small repulsion ($\varepsilon_{\mathrm{pp}} \lesssim 1$),
two types of bound membranes are mixed (see Fig.~\ref{fig:symr}(a)).
Since the spontaneous curvature of particle pairs of different types of the bound membrane sites (u and d) is zero,
more contact between these sites results in lower bending energy in a flat membrane.
At stronger repulsion  ($\varepsilon_{\mathrm{pp}} \gtrsim 2$),
the membrane forms larger domains (larger squares and wider stripes for $\mu_{\mathrm{u}}/k_{\mathrm{B}}T = 7.5$ and $20$, respectively) 
and has a larger vertical span $z_{\mathrm {mb}}$  (see Fig.~\ref{fig:symr}(b) and (d)).
At $\mu_{\mathrm{u}}/k_{\mathrm{B}}T = 7.5$ and $\varepsilon_{\mathrm{pp}} \geq 2.5$,
the domain size has a wide distribution owing to the hysteresis (see the large error bars in Fig.~\ref{fig:symr}(d)). At $\varepsilon_{\mathrm{pp}} = 2$,
this hysteresis is relatively weak, and
the membrane continuously changes the domain structures according to the variation in the chemical potentials (see Movies S1 and S2 in ESI from $\mu_{\mathrm{u}}/k_{\mathrm{B}}T = 6$ to $7.5$ and from $\mu_{\mathrm{u}}/k_{\mathrm{B}}T = 7.5$ to $12$, respectively).

Next, we describe the case that the up--down symmetry is broken in thermal equilibrium,
i.e., $\mu_{\mathrm{u}} \ne \mu_{\mathrm{d}}$ and $\mu_{\mathrm{ff}}= \mu_{\mathrm{d}}-\mu_{\mathrm{u}}$.
When the lipid composition in two leaflets of a bilayer membrane is different,
this asymmetric condition can be generated.
With increasing $\mu_{\mathrm{ff}}$ at $\mu_{\mathrm{d}}/k_{\mathrm{B}}T = 7.5$, 
the domain of the upper proteins becomes smaller with respect to the domains of the lower proteins,
and  one lower domain is in contact with five or more upper domains.
The angles of the contact points become more and less than $90^{\circ}$ for lower and upper domains, respectively.
Consequently, at $\mu_{\mathrm{ff}}/k_{\mathrm{B}}T = 1$, the membrane form a kagome lattice; 
the hexagonal domain of the lower proteins is surrounded by six triangular domains of the upper proteins (see Fig.~\ref{fig:cp75asym}(a)).
At $\mu_{\mathrm{ff}}/k_{\mathrm{B}}T \gtrsim 1.5$, this lattice is broken so that the lower domain is percolated (see Fig.~\ref{fig:cp75asym}(b) and (d)).
For a stronger repulsion ($\varepsilon_{\mathrm {pp}}=3$), the membrane forms polygonal lattices and
the coexistence of triangular and square domains also occurs (see Fig.~\ref{fig:cp75ar3}).
For $\mu_{\mathrm{d}}/k_{\mathrm{B}}T = 20$, similarly, with increasing $\mu_{\mathrm{ff}}$,
the striped domains of the upper proteins are separated into thread-like domains,
and subsequently, small circular domains are formed (see Fig.~\ref{fig:cp20asym}).
The length of the thread-like domains frequently changes via the domain fusion and division.

After the equilibration, 
the binding/unbinding or flip--flop processes or both can be stopped.
The phase behavior is maintained except for the density fluctuations owing to the detailed balance.
Thus, the aforementioned domains are formed even in the absence of the binding/unbinding and/or flip--flop processes.
Thus, by adjusting the protein densities, transmembrane proteins with only lateral diffusion (no unbinding and no flip--flop) can form the checkerboard domains.

\begin{figure}[t]
\includegraphics[]{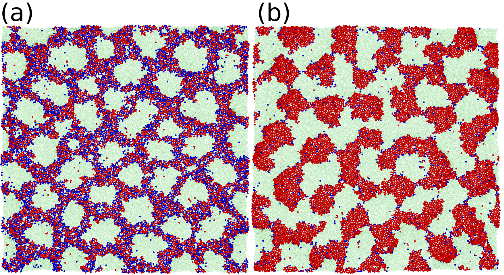}
\includegraphics[]{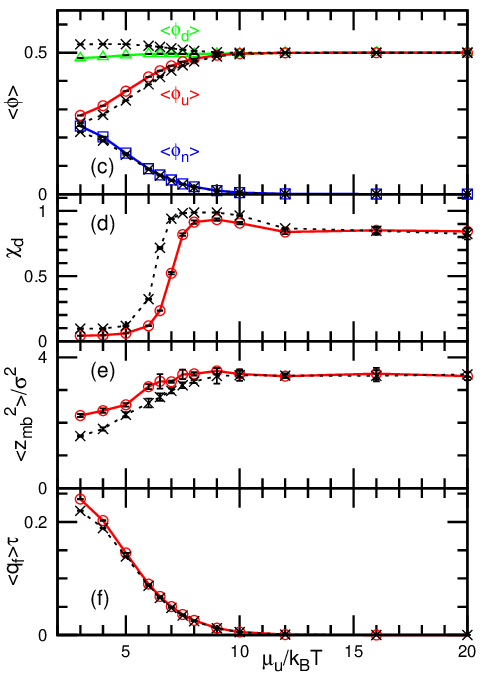}
\caption{
Dependence on the binding chemical potential $\mu_{\mathrm{u}}$ for the upper surface 
at $\mu_{\mathrm{d}}/k_{\mathrm{B}}T=20$,  $\mu_{\mathrm{ff}}=0$, and $\varepsilon_{\mathrm {pp}}=2$. 
(a), (b) Snapshots for (a) $\mu_{\mathrm{u}}/k_{\mathrm{B}}T=3$ and (b)  $\mu_{\mathrm{u}}/k_{\mathrm{B}}T=7$.
(c) Densities of upper proteins $\phi_{\mathrm{u}}$, lower proteins $\phi_{\mathrm {d}}$,
and unbound membrane $\phi_{\mathrm {n}}$.
(d) Normalized mean cluster sizes $\chi_{\mathrm{d}}$ for the lower proteins.
(e) Membrane vertical span $z_{\mathrm {mb}}$.
(f) Flow rate $q_{\mathrm{f}}$ of proteins from the lower to upper solutions.
The solid and dashed lines represent the data 
for the normal and ten-fold slower flip--flop rates, respectively.
}
\label{fig:cp20neq}
\end{figure}

\begin{figure}[t]
\includegraphics[]{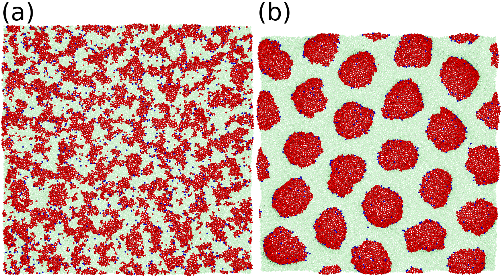}
\includegraphics[]{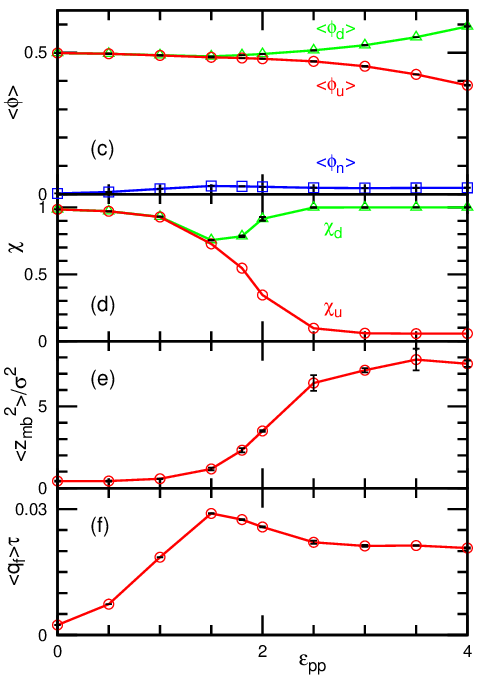}
\caption{
Dependence on the repulsion strength $\varepsilon_{\mathrm {pp}}$ 
at $\mu_{\mathrm{d}}/k_{\mathrm{B}}T=20$,  $\mu_{\mathrm{u}}/k_{\mathrm{B}}T=8$, and  $\mu_{\mathrm{ff}}=0$.
(a), (b) Snapshots for (a) $\varepsilon_{\mathrm {pp}}=1.5$ and (b)  $\varepsilon_{\mathrm {pp}}=3$.
(c) Densities of upper proteins $\phi_{\mathrm{u}}$, lower proteins $\phi_{\mathrm {d}}$,
and unbound membrane $\phi_{\mathrm {n}}$.
(d) Normalized mean cluster sizes $\chi_{\mathrm{u}}$ and $\chi_{\mathrm{d}}$.
(e) Membrane vertical span $z_{\mathrm {mb}}$.
(f) Flow rate $q_{\mathrm{f}}$ of proteins from the lower to upper solutions.
}
\label{fig:cp20d8r}
\end{figure}

\begin{figure}[t]
\includegraphics[]{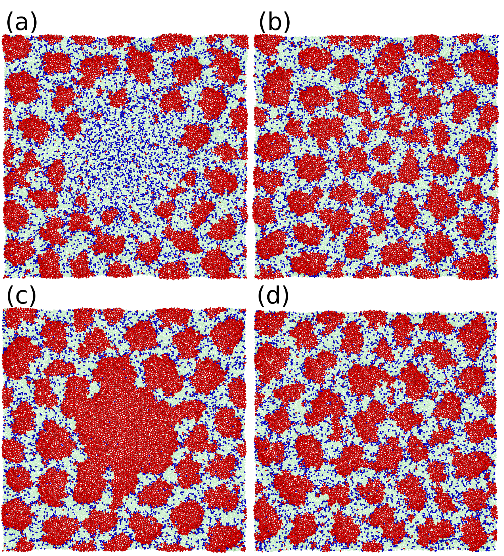}
\includegraphics[]{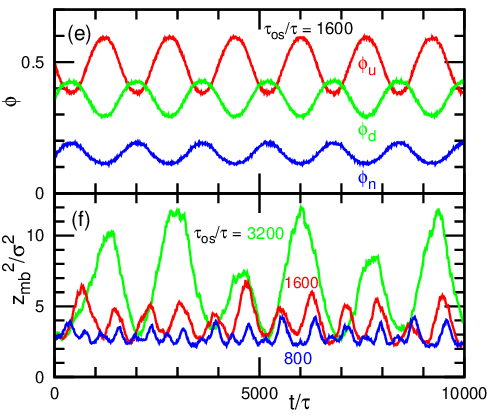}
\caption{
Time evolution of the membrane under an oscillatory change in $\mu_{\mathrm{ff}}$
at  $\mu_{\mathrm{os}}/k_{\mathrm{B}}T=4$, $\sigma_{\mathrm{os}}=30\sigma$, $\mu_{\mathrm{u}}/k_{\mathrm{B}}T=20$, $\mu_{\mathrm{d}}/k_{\mathrm{B}}T=5$, and $\varepsilon_{\mathrm {pp}}=2$. 
(a)--(d) Sequential snapshots at (a) $t/\tau=5200$, (b) $t/\tau=5600$, (c) $t/\tau=6000$, and (d) $t/\tau=6400$ for $\tau_{\mathrm {os}}/\tau=1600$
($t= \tau_{\mathrm{os}}(3+0.25n)$ with $n=1$, $2$, $3$, and $4$, respectively).
(e) Time development of the densities at $\tau_{\mathrm{os}}/\tau=1600$.
(f) Time development of the membrane vertical span $z_{\mathrm {mb}}$ at $\tau_{\mathrm{os}}/\tau=800$, $1600$, and $3200$.
}
\label{fig:e4w30}
\end{figure}

\subsection{Phase Separation in Nonequilibrium}\label{sec:neq}

In this subsection, we describe the steady states in a nonequilibrium ($\mu_{\mathrm{ff}} \ne \mu_{\mathrm{d}}-\mu_{\mathrm{u}}$).
First, we consider that the upper and lower solutions and two leaflets of the membrane are identical ($\mu_{\mathrm{d}}=\mu_{\mathrm{u}}$), 
but the flip--flop is asymmetric ($\mu_{\mathrm{ff}}\ne 0$). The flip process is activated by an enzyme with an energy consumption such as ATP hydrolysis, 
and represented by a positive value of $\mu_{\mathrm{ff}}$.
In this case, the phase behavior is close to that in the thermal equilibrium with the same value of $\mu_{\mathrm{ff}}$,
as shown in Figs.~\ref{fig:cp75asym} and \ref{fig:cp20asym}.
Since the protein unbinding is very slow at $\mu_{\mathrm{u}}/k_{\mathrm{B}}T=7.5$, the membrane domains are dominantly determined by the flip--flop
and only slight differences are observed, as shown in  Fig.~\ref{fig:cp75asym}.
However, the proteins flow from the upper to lower solutions through this slow unbinding of the lower surface (see Fig.~\ref{fig:cp75asym}(f)).
In contrast, at $\mu_{\mathrm{u}}/k_{\mathrm{B}}T=20$,
the unbinding is negligibly slow so that
  $\phi_{\mathrm {n}} \leq 1\times 10^{-6}$, $|\langle q_{\mathrm{f}}\rangle|\tau \leq 1\times 10^{-6}$, and
 the dashed lines in Fig.~\ref{fig:cp20asym} are completely overlapped with the solid lines.

Next, we consider the case in which the binding chemical potentials onto the upper and lower surfaces are different,
i.e.,  $\mu_{\mathrm{d}} \ne \mu_{\mathrm{u}}$, while the flip--flop is symmetric, $\mu_{\mathrm{ff}}=0$ (Figs.~\ref{fig:cp75asym}, \ref{fig:cp20neq}, and \ref{fig:cp20d8r}).
This condition is easily constructed during the experiments by changing the protein concentrations in the upper and lower solutions
(for a dilute solution, $\mu=\mu_0 + k_{\mathrm{B}}T\ln(\rho)$, where $\rho$ is the protein concentration in bulk).
We vary $\mu_{\mathrm{u}}$ at  $\mu_{\mathrm{d}}/k_{\mathrm{B}}T=7.5$
(see cyan lines in Fig.~\ref{fig:cp75asym}) and at $\mu_{\mathrm{d}}/k_{\mathrm{B}}T=20$ (see Fig.~\ref{fig:cp20neq}).
As the difference, $\Delta\mu = \mu_{\mathrm{d}}-\mu_{\mathrm{u}}$, increases,
the density $\phi_{\mathrm {n}}$ of unbound membranes increases,
whereas $\phi_{\mathrm {d}}$ is almost constant (see Figs.~\ref{fig:cp75asym}(c) and \ref{fig:cp20neq}(c)).
At $\mu_{\mathrm{d}}/k_{\mathrm{B}}T=20$, the domains of the lower proteins transform from stripes to circular shapes (Fig.~\ref{fig:cp20neq}(a)). 
This is a continuous transition characterized by the change $\chi_{\mathrm {d}}\simeq 1$ to $0$, as shown in Fig.~\ref{fig:cp20neq}(d).
The flow from the lower to upper solution increases to reflect the $\phi_{\mathrm {n}}$ increase (compare Fig.~\ref{fig:cp75asym}(c) with (f) and also Fig.~\ref{fig:cp20neq}(c) with (f)).
It should be noted that this is the opposite flow with the same amplitude to that at $\mu_{\mathrm{ff}}\ne 0$ and $\mu_{\mathrm{d}}=\mu_{\mathrm{u}}$
in the linear nonequilibrium conditions (see the overlap of two lines at $\Delta\mu/k_{\mathrm{B}}T\leq 1$ in Fig.~\ref{fig:cp75asym}(f)).
Figure~\ref{fig:cp20d8r} shows the dependence on $\varepsilon_{\mathrm {pp}}$ at $\mu_{\mathrm{d}}/k_{\mathrm{B}}T=20$ and $\mu_{\mathrm{u}}/k_{\mathrm{B}}T=8$.
At a strong repulsion ($\varepsilon_{\mathrm {pp}} \geq 3$), the upper proteins form circular domains and the lower proteins form hexagonal lattices (see Fig.~\ref{fig:cp20d8r}(b)).
The flow rate $q_{\mathrm {f}}$ and $\phi_{\mathrm {n}}$ have maxima at $\varepsilon_{\mathrm {pp}}=1.5$ (see Fig.~\ref{fig:cp20d8r}(c) and (f)), 
in which the membrane starts forming bumped domains (see Fig.~\ref{fig:cp20d8r}(e)).

Finally, we consider a stimulation by an oscillatory change in chemical potentials.
The flip--flop chemical potential is temporally varied in a local region as 
\begin{equation}
\mu_{\mathrm{ff}}(r,t) = \mu_{\mathrm{os}}\sin(2\pi t/\tau_{\mathrm{os}})\exp(- r^2/2{\sigma_{\mathrm{os}}}^2 ),
\end{equation}
where $r= \sqrt{x^2+y^2}$ is the lateral distance from the center of the simulation box.
Figure~\ref{fig:e4w30} shows the membrane dynamics at $\mu_{\mathrm{os}}/k_{\mathrm{B}}T=4$ and $\sigma_{\mathrm{os}}=30\sigma$.
The stimulated region (the circular region at the center) 
exhibits the oscillatory formation of the upper protein domains, but the other region remains in the steady state
(see Fig.~\ref{fig:e4w30}(a)--(d) and corresponding movie (Movie S3) in ESI).
The total protein densities and membrane vertical span oscillate according to the local domain formation (see Fig.~\ref{fig:e4w30}(e) and (f), respectively).
For slower oscillations (longer $\tau_{\mathrm{os}}$), the membrane exhibits larger deformation, since the membrane shape can be further relaxed.

We also examined the membrane behavior under the different conditions (e.g, checkerboard-shaped and striped domains), including oscillation of $\mu_{\mathrm{u}}$.
The oscillations occur locally between the domain-structures expected from the chemical potentials. However, the membrane far from the stimulated region similarly remains in its steady state.
Thus, traveling waves are not generated in the present system.

\begin{figure}[t]
\includegraphics[]{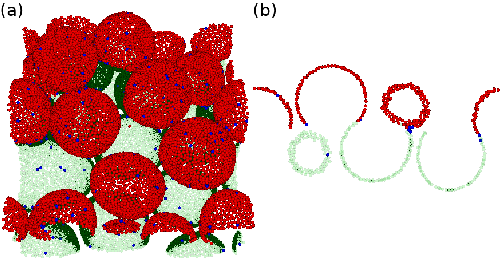}
\caption{
Snapshots of a tensionless membrane ($\gamma=0$) under an up--down symmetrical condition at $\mu_{\mathrm{u}}/k_{\mathrm{B}}T=\mu_{\mathrm{d}}/k_{\mathrm{B}}T=7.5$, $\mu_{\mathrm{ff}}=0$, and $\varepsilon_{\mathrm {pp}}=2$. (a) A bird's-eye view. (b) A sliced membrane from a side view.
}
\label{fig:snapt0}
\end{figure}

\begin{figure}[t]
\includegraphics[]{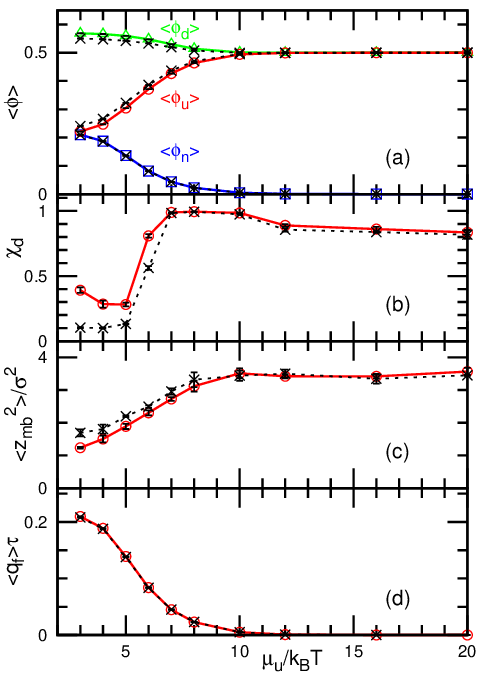}
\caption{
Protein binding with a ten-fold faster binding/unbinding rate ($0.1$ trials per particle) for
the condition used in Fig.~\ref{fig:cp20neq}.
(a) Densities of upper proteins $\phi_{\mathrm{u}}$, lower proteins $\phi_{\mathrm {d}}$,
and unbound membrane $\phi_{\mathrm {n}}$.
(b) Normalized mean cluster sizes $\chi_{\mathrm{d}}$ for the lower proteins.
(c) Membrane vertical span $z_{\mathrm {mb}}$.
(d) Flow rate $q_{\mathrm{f}}$ of proteins from the lower to upper solutions.
The solid and dashed lines represent the data, in which 
the  flip--flop and binding/unbinding MC processes are first performed, respectively.
}
\label{fig:cp20a10}
\end{figure}

\section{Summary}\label{sec:sum}

We have studied the phase behavior of membranes induced by the binding/unbinding of curvature-inducing molecules such as proteins and conical surfactants.
The proteins (or other amphiphilic molecules) can bind the upper and lower surfaces of the membrane and move between two surfaces via the flip--flop.
In up--down symmetric conditions, the proteins bind onto the two surfaces with the same amount.
The protein domains exhibit checkerboard patterns as well as spot and stripe patterns.
The vertices of the checkerboard consist of the unbound membranes, and the square regions bend alternatively in the upward and downward directions.
In asymmetric conditions, the kagome lattice (a lower hexagonal domain surrounded by six triangular domains of the upper proteins)
and thread-like domains are formed.
In nonequilibrium steady states, the flow of proteins between the upper and lower solutions occurs via the flip--flop.

In nonequilibrium,
membranes can exhibit characteristic patterns and dynamics.
In living cells, protein binding  can be activated or inactivated by the consumption of ATP/GTP.
In addition, proteins, such as flippases, can transport molecules between two leaflets of bilayers.
Here, we investigate  a simple model system to understand biological membranes under the flows of molecules through membranes.
In theoretical studies,
energy-independent binding/unbinding have often been used as nonequilibrium processes\cite{gout21,rao01,rama15}
However, experimental situations are typically complicated, since these processes are accompanied by conformational changes of proteins.
In the present system, a nonequilibrium condition can be generated by the chemical-potential difference between upper and lower solutions, which is easily set up in experiments.
Experimentally, flows of small surfactant molecules from outer to inner solutions of a vesicle
have been obtained via binding, unbinding, and flip--flop.\cite{miel20,holl21}
They can induce a shape transition of the vesicle in tens of seconds.
Although the membrane is homogeneous in these experiments,
phase separation likely occurs by using two types of surfactants, in which hydrophobic segments consist of alkyl or fluoroalkyl chains, 
 since these chains can exhibit phase separation.\cite{xiao16,waka22}
Thus, the present simulation results can be experimentally examined.

In this study, we obtained only steady states in nonequilibrium.
The local stimulation by the oscillation of chemical potentials does not modify the phase behavior far from the stimulation region.
Previously, traveling waves were reported in other systems both in one-dimensional
\cite{cagn18} and two-dimensional geometries.\cite{zaki18} The traveling waves have also been obtained by the reaction-diffusion dynamics coupled with membrane deformation.\cite{pele11,wu18,tame21,tame22} The addition of different types of interactions to the present system may lead to more complex dynamics such as traveling waves. This problem can be resolved as part of future research.

\begin{acknowledgments}
This work was supported by JSPS KAKENHI Grant Number JP21K03481.
\end{acknowledgments}

\begin{appendix}

\section{Structure of tensionless membranes}\label{app1}

Here, we briefly describe the membrane structures at $\gamma=0$.
When the surface tension is reduced to $\gamma=0$, the membrane laterally shrinks
and spherical buds are formed at up--down symmetrical conditions. 
Owing to high curvatures, 
membrane is ruptured at the boundary of unbound domains
and pores open at the vertices of the checkerboard pattern as shown in Fig.~\ref{fig:snapt0}.
This periodic spherical-cap pattern is also formed at  $\mu_{\mathrm{u}}/k_{\mathrm{B}}T=\mu_{\mathrm{d}}/k_{\mathrm{B}}T=20$,
whereas  stripe domains are stable at $\gamma=0.5k_{\mathrm{B}}T/\sigma^2$.

\section{Dependence on binding/unbinding and flip/flop rates}\label{app2}

In this study, we used one MC trial per particle for flip--flop and $0.01$ trial per particle for binding/unbinding
every MC step (time interval of $\tau_{\mathrm {MC}}$).
Here, we discuss the dependence on these rates in nonequilibrium steady states.
When the flip--flop rate is set ten-fold slower ($0.1$ MC trial per particle),
the densities of the lower and upper proteins slightly increase and decrease, respectively, 
at $\mu_{\mathrm{u}}/k_{\mathrm{B}}T \lesssim 10$.
Consequently, the percolation of lower-protein domains remains at the slightly lower value of $\mu_{\mathrm{u}}$
and the membrane is slightly flatter than at the normal flip--flop rate (compare solid and dashed lines in Fig.~\ref{fig:cp20neq}).
The flow rate $q_{\mathrm{f}}$ is slightly reduced at  $\mu_{\mathrm{u}}/k_{\mathrm{B}}T \lesssim 4$.

When the binding/unbinding rate is set ten-fold faster ($0.1$ trial per particle),
similar dependence is obtained (see Fig.~\ref{fig:cp20a10}).
However, the results slightly depend on the details of the MC processes in this condition.
In our simulation code, at every MC step, all flip--flop MC processes are performed, 
and later all binding/unbinding processes are performed.
When the opposite sequence is used (dashed lines in Fig.~\ref{fig:cp20a10}), 
the density changes in the upper and lower proteins slightly decrease.
When a hundred-fold faster binding/unbinding is used (one trial per particle),
these reductions are enhanced.
This is because the latter process has a stronger effect far from thermal equilibrium.
The flip--flop reduces the density difference between $\phi_{\mathrm {u}}$ and  $\phi_{\mathrm {d}}$  at $\mu_{\mathrm{ff}}=0$.
To avoid dependence on this code detail, we chose the present MC rates.
Using this rate, the code dependence is undetectably small in all simulation conditions in this study.

\end{appendix}

\end{document}